# Classification of ST and Q Type MI variant using thresholding and neighbourhood estimation method after cross wavelet based analysis

**Swati Banerjee[1], M. Mitra[1]**

[1] Department of Applied Physics, Faculty of Technology, University of Calcutta

92, APC Road, Kolkata-700009, India

swatibanerjee@ieee.org, mmitra@ieee.org

*Abstract*—

Most of the ECG analysing systems use explicit time plane features for cardiac pattern classification. This paper proposes a cross wavelet transform based method for Electrocardiogram signal analysis where parameters are identified from wavelet cross spectrum and wavelet cross coherence of ECG patterns. Application of this technique for classification eliminates the need for extraction of various time plane features. The cross-correlation is the measure of similarity between two waveforms. Application of the Continuous Wavelet Transform to two time series and the cross examination of the two decomposition reveals localized similarities in time and scale. Parameters extracted from Wavelet Cross Spectrum (WCS) and Wavelet Coherence (WCOH) is used for classification. A pathologically varying pattern in QT zone of inferior lead III shows the presence of Inferior Myocardial Infarction (IMI). The Cross Wavelet Transform and Wavelet Coherence is used for the cross examination of single normal and abnormal (IMI) beats. A normal template beat is selected as the absolute normal pattern. Computation of the WCS and WCOH of normal template and various other normal and abnormal beat reveals the existence of variation among patterns under study. The Wavelet cross spectrum and Wavelet coherence of various ECG patterns shows distinguishing characteristics over two specific regions R1 and R2, where R1 is the QRS complex area and R2 is the T wave region. Parameters are identified for classification of Type 1 IMI (non Q type, with ST elevation and attenuated QRS complex) and Type 2 IMI (Q type MI with deep Q and inverted T) and normal subjects. Accuracy of the proposed classification method is obtained as 99.43% for normal and abnormal class.

*Keywords- Cross wavelet transform, Wavelet Spectrum, Myocardial Infarction, Interpolation, Electrocardiogram.*

1. Introduction

Electrocardiogram signals are intrinsically non-stationary because their underlying statistical properties changes with time. This source of non-stationarity is intrinsic in nature as the origins are due to physiological changes. This makes wavelet transform an effective tool for analysis of biomedical signals. Coronary heart disease, also called coronary artery disease is one of the predominant health concerns all over the world. Electrocardiogram is the interpretation of the electrical activity of the heart over a period of time [1], which is an effective low cost diagnostic tool for screening of cardiac abnormalities. Good performance of any automatic ECG analysing system depends upon the reliable and accurate detection of the basic features. QRS detection is necessary to determine the heart rate and is used as reference for beat alignment. The automatic delineation of ECG is widely studied and many algorithms are developed for identification of QRS complex and other clinical signatures [2]-[4]. Wavelet transforms have been applied to ECG signals for enhancing late potentials [5], reducing noise [6], QRS detection [7], normal and abnormal beat recognition [8] and delineation of ECG characteristic features [11]. The methods used in these studies were conducted through continuous wavelet transform [9], multiresolution analysis [8, 9] and dyadic wavelet transform [10]. Many classification methods each with distinguishing characteristics have been developed using neuro- fuzzy [12, 13] and self organizing maps [14, 15, 16]. A rule mining based method is developed in [17], where ischemic beats are identified by extraction of features followed by feature discretization and rule mining. ECG features are also extracted using linear predictive coding in [18].

In this work a method for analysis of ECG data by the method of cross-wavelet transform (XWT) is proposed. Before any form of analysis the signal is denoised and R peaks are registered. The heart rate is computed and then beats are segmented for analysis. Each of the segmented beats is time normalized before analysis because the heart rate varies from subject to subject. For this study only Inferior MI (IMI) and normal class is considered. IMI is identifiable from the inferior leads II, III, aVF, of which lead III is selected for the present analysis. All the input data for this method has been selected from Physikalisch-Technische Bundesanstalt diagnostic ECG database (ptb-db) [20]. A pathologically normal patient is selected as standard normal and an extracted beat is labeled, as the standard normal template beat. Normal and abnormal ECG patterns are analyzed by subjecting them to XWT. The wavelet cross spectrum and wavelet coherence reveals various distinguishing characteristics when compared to normal and abnormal ECG data. A heuristically determined mathematical formula extracts parameter from the WCS and WCOH. Empirical tests establish that these parameters are relevant for classification of normal and abnormal Cardiac patterns. In this work a threshold based classifier is proposed because the extracted parameters have sufficiently distinct class separability characteristic. Once the parameters are determined threshold value based normal and abnormal classification is accomplished. Once the normal and abnormal classes are identified we further proceed on to distinguish between Type 1 IMI (non Q type, with ST elevation and attenuated QRS complex) and Type 2 IMI (Q type MI with deep Q and inverted T). The problem then reduces to a hierarchical classification problem where there exists two broad category of classes normal and abnormal (IMI) and again IMI class consists of Type 1 and Type 2 class. Threshold based estimation fails in identification of the Type 1 and Type 2 IMI and for this k-nn rule is employed.

## 2. Theory

### 2.1 Continuous wavelet transform(CWT)

The continuous wavelet transform involves decomposing a signal *f(t)*, into a number of translated and dilated wavelets. The main idea behind this is to take a *mother* wavelet $\psi(t)$, translate and dilate it, convolve it with the function of interest, and map out the coefficients in *wavelet space*, spanned by translation and dilation. Periodic behavior, then shows up as a pattern spanning all translations at a given dilation, and this redundancy in the wavelet space makes detection of periodic behavior rather easy. The wavelet transform preserves temporal locality which is an advantage over Fourier analysis.

### 2.2 Cross Wavelet transform (XWT) and Wavelet Coherence (WCOH)

The cross wavelet transform (XWT) of two time series $x_n$ and $y_n$, is defined as

$$W^{XY} = W^X W^{Y*} \qquad (1)$$

Where * denotes complex conjugation. We further define the cross wavelet power as $|W^{XY}|$. The complex argument $\arg(W^{XY})$ can be interpreted as the local relative phase between $x_n$ and $y_n$ in time frequency space. The theoretical distribution of the cross wavelet power of two time series with background power spectra $P_k^X$ and $P_k^Y$ is given by Torrence and compo in [21, 22]. Another useful measure is how coherent the cross wavelet transform is in time frequency space. Following, Torrence and Webster [22], the wavelet coherence of two time series is defined as:

$$R_n^2(s) = \frac{\left|S(s^{-1}W_n^{XY}(s))\right|^2}{S(s^{-1}|W_n^X(s)|^2) \cdot S(s^{-1}|W_n^Y(s)|^2)} \qquad (2)$$

Where, S is a smoothing operator. And wavelet coherence can be thought of as a localized correlation coefficient in time frequency space.

## 3. Material and Method

The proposed methodology consists of denoising of ECG data followed by R peak registration and beat segmentation. The heart rate is a variable quantity and accordingly the beat duration changes. So, each of the segmented beat is time normalized. These beats are subjected to further analysis. The cross wavelet analysis reveals much significant information about the characteristic of the ECG beats under study. Features are determined from the generated Wavelet Cross

Spectrum (WCS) and Wavelet Coherence (WCOH) which are further empirically validated. The detailed description of the method is illustrated by Fig. 1.

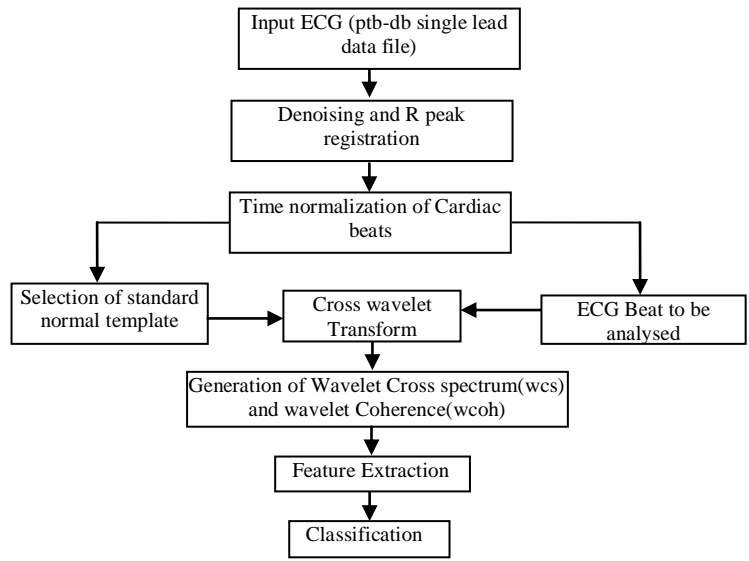

**Fig1:** Flow chart for the analysis

*3.1    Data*

For the analysis all the ECG data is taken from ptb-db diagnostic database. One cardiac beat from a 25 year old, pathologically normal, male subject with a heart rate of 72 beats/min, is considered as the standard normal template for analysis. Notable, morphological differences exist for normal Inferior MI subjects in the QT zone of the inferior leads. Inferior lead III is used to show the results of the analysis.

*3.2    R Peak Registration*

Denoising of ECG data is an essential step before any form of analysis as this increases the efficiency of the algorithm. Present work uses DWT based decomposition and selective reconstructions of wavelet coefficients for denoising and QRS detection. The denoising and basic R peak registration technique used for this work is the method developed in [11].

*3.3    Time normalization of cardiac cycles*

Once the R peaks are registered, the R-R interval is computed and divided into 1:2 ratios (Say, x: 2x points). One cardiac cycle gives the details of the pathological condition of the patient and hence each beat needs to be segmented before subjecting it to cross-wavelet analysis.  Considering x points to the left and 2x points to the right of R index one cardiac beat is extracted. FFT based interpolation technique [19] is used for time normalization of each beat segment, as the heart rate varies for each subject. In this study all beats are normalized to 1000 samples. The time normalization is important for comparability of two different patterns and finding out notable differences and variations in the same time scale.

### 3.4 Wavelet Cross Spectrum (WCS) and Wavelet Coherence (WCOH) analysis of ECG beats

The application of CWT to two time series and the cross examination of the two decompositions reveals localized similarities in time and scale (scale being nearly inverse of frequency). The WCS and WCOH are used for cross examination of a single normal and abnormal (IMI) beat with that of a standard normal template beat. Because of the morphological similarity with that of the QRS complex, morlet is selected as the mother wavelet for analysis. Similarity between the two signals under study can be estimated by analysis of the spectral pattern generated by WCS and the coherence by WCOH. In this analysis 512 scales are used.

Fig. 2 and 3, shows the distinguishing regions R1, the QRS complex region and R2, the T wave region. It is evident from the color coded plots, that there exist distinct variations in the spectral and coherence components, revealing the nature of the analysed signals. A heuristically determined mathematical formula extracts parameter from the WCS and WCOH. Empirical tests establish that these parameters are relevant for classification of normal and abnormal Cardiac patterns. Fig. 2(a, b, c) shows Type 1 IMI (non Q type, with ST elevation and attenuated QRS complex) and Fig.2 (d, e, f) Type 2 IMI (Q type MI with deep Q and inverted T) where as fig. 3 shows normal cases respectively. Also it is evident from Fig. 2(a, b, c) and Fig. 2(d, e, f) that the Type 1 and Type 2 IMI produces visibly different values of WCS and WCOH. Once the appropriate parameters are extracted a further classification of Type 1 and Type 2 IMI along with the normal class is possible. The extraction of parameters from the generated WCS and WCOH is discussed in the next section.

### 3.5 Feature Extraction and parameter identification

The XWT generates WCS and WCOH, which are matrices containing the wavelet cross spectrum and wavelet coherence between two signals. In this study normal subjects and subjects with IMI is considered. Distinction between Type I and Type II MI is accomplished using nearest neighbor based classification rule. All the results are analyses over Lead III. There exists a pathologically varying QT pattern for normal and abnormal data. The QT zone being the pathological region is selected for parameter extraction and analysis. A span of 80 points from the left to 400 points to the right of the registered R peak is the QT zone [1, 2]. The start and end of the time zone is marked as *t1* and *t2* respectively. Visual inspection of the colour coded spectrogram for WCS and WCOH in fig.2, 3 reveals that, the effect of the analysis is prominent over a particular scale range. Heuristically determined mathematical formulas are developed and tested over several data sets. These equations are for feature extraction from WCS and WCOH.

*Scale selection from Wavelet Cross Spectrum*

To find the actual span of scales that are significant for analysis, a time-scale relation is established. Equation 3 is developed from the WCS, where *s*, signifies the scale and varies from 1 to 512.

$$sum\_WCS(s) = \sum_{t1}^{t2} WCS(s,t) \qquad (3)$$

The variable **sum_WCS(s)** contains the summation of the WCS value at each scale over the whole ECG beat. The **sum_WCS(s)** values of several subjects both normal and abnormal are plotted in fig.4. By inspecting the graph it is found that variation of the cross wavelet spectrum for normal and IMI subject is most prominent in the scale range marked with **s1** to **s2**. The normal and abnormal data plot in fig.4 shows that visible differences exists in the scale range of s1=75 and s2=300, showing a significant positive and negative value respectively corresponding to the pathological beats where the normal is plotted with the solid lines and abnormal with that of the dotted line. Any parameter (or feature) extracted from this scale range, over the QT zone from WCS and WCOH will produce a unique identification signature which can be used for classification of normal and abnormal cardiac patterns.

*Wavelet Cross Spectrum (WCS) and Wavelet Coherence (WCOH) based Parameter extraction*

The parameters extracted from the WCS and WCOH matrices over the selected scale range of *s1* to *s2* and over the QT segment *t1* to *t2*.

So, eq.4 and eq.5 is framed for parameter extraction which is used for classification.

$$sum\_WCOH\_scale(pa) = \sum_{s1}^{s2} \sum_{t1}^{t2} WCOH(s,t) \qquad (4)$$

$$sum\_WCS\_scale(pb) = \sum_{s1}^{s2} \sum_{t1}^{t2} WCS(s,t) \qquad (5)$$

Accordingly, threshold values (*th1* and *th2*) from eq.4 and eq.5 are estimated for normal and abnormal class identification. Set of parameter (pa, pb) are extracted from the standard leads III. From the extracted parameters and also from the visual inspection of the spectrograms of Fig. 3 and 4 there exists variation in the WCS and WCOH values. Also form eq. 4 and 5 identification of Type 1 and Type 2 IMI becomes evident.

### 3.6 Classification

In the present scenario the data classification reduces to a two class problem creating a partition between normal and IMI subject. In this work a threshold based classifier is used because the extracted parameters have sufficiently distinct class separability characteristic. Once the parameters are determined, threshold values *(th11 and th12)* for *pb* and *pa* respectively is estimated to distinguish different classes. Once the normal and abnormal classes are identified we further proceed on to distinguish the Type 1 IMI (non Q type, with ST elevation and attenuated QRS complex) and Type 2 IMI (Q type MI with deep Q and inverted T). This problem then reduces to a hierarchical classification problem where there exists two broad category of classes normal and abnormal (IMI) and again IMI class consists of Type 1 and Type 2 class. Threshold based estimation fails in identification of the Type 1 and Type 2 IMI and for this k-nn rule is employed.

The generated *threshold based classification rule* for identification of normal and IMI class is stated below:

- If *pa1<th11 and pb1<th12*

    then 'abnormal' mark as '0'

    else 'normal' mark as '1'

    end if

**The Nearest Neighbor Rule based Classification** for identification of Type 1 and Type 2 IMI is illustrated below:

Nearest neighbor (NN) is one of the most popular classification rules, although it is an old technique. Here $c$ classes exists, $\omega_i$, $i = 1, 2, \ldots, c$, and a point $x \in R^l$, and $N$ *training* points, $x_i$, $i = 1, 2, \ldots, N$, in the $l$-dimensional space, with the corresponding class labels. Given a point, $x$, whose class label is unknown, the task is to classify $x$ in one of the $c$ classes. The rule consists of the following steps:

**1.** Among the $N$ training points, search for the $k$ neighbors closest to $x$ using a distance measure. The parameter $k$ is user-defined.

**2.** Out of the $k$-closest neighbors, identify the number $k_i$ of the points that belong to class $\omega_i$. Obviously,

$$\sum_{i=1}^{c} k_i = k$$

**3.** Assign $x$ to class $\omega_i$, for which $ki > kj$, $j \neq i$. In other words, $x$ is assigned to the class in which the majority of the $k$-closest neighbors belong.

Section 4 shows the results and empirical validation of the extracted parameter.

## *4. Results*

All the input data for this method has been selected from ptb-db of Physionet [23], which contains 549 records from 290 subjects with 52 healthy controls and 148 Myocardiac infarction patients. A normal non smoker male subject of 25 years old, is considered as the normal template for analysis, the patient id is: ptbdb/patient150/s0287lre.

## 4.2 Analysis of morphologically varying ECG data by WCS and WCOH

The below shown figures reveals differences in the WCS and WCOH, when normal-normal and normal-abnormal pairs are subjected to XWT analysis. Two major region of difference is marked as R1 and R2. Where R1 is the QRS Complex region and R2 is the T wave region. Fig. 2(a, b, c) shows Type 1 IMI (non Q type, with ST elevation and attenuated QRS complex) and Fig.2 (d, e, f) Type 2 IMI (Q type MI with deep Q and inverted T) where as Fig. 3 shows normal cases. From the colour coded spectrogram for WCS and WCOH it is evident that spectral and coherence variations exist in region R1 and R2. For concise presentation of data and due to space constrains limited number of analysis is shown.

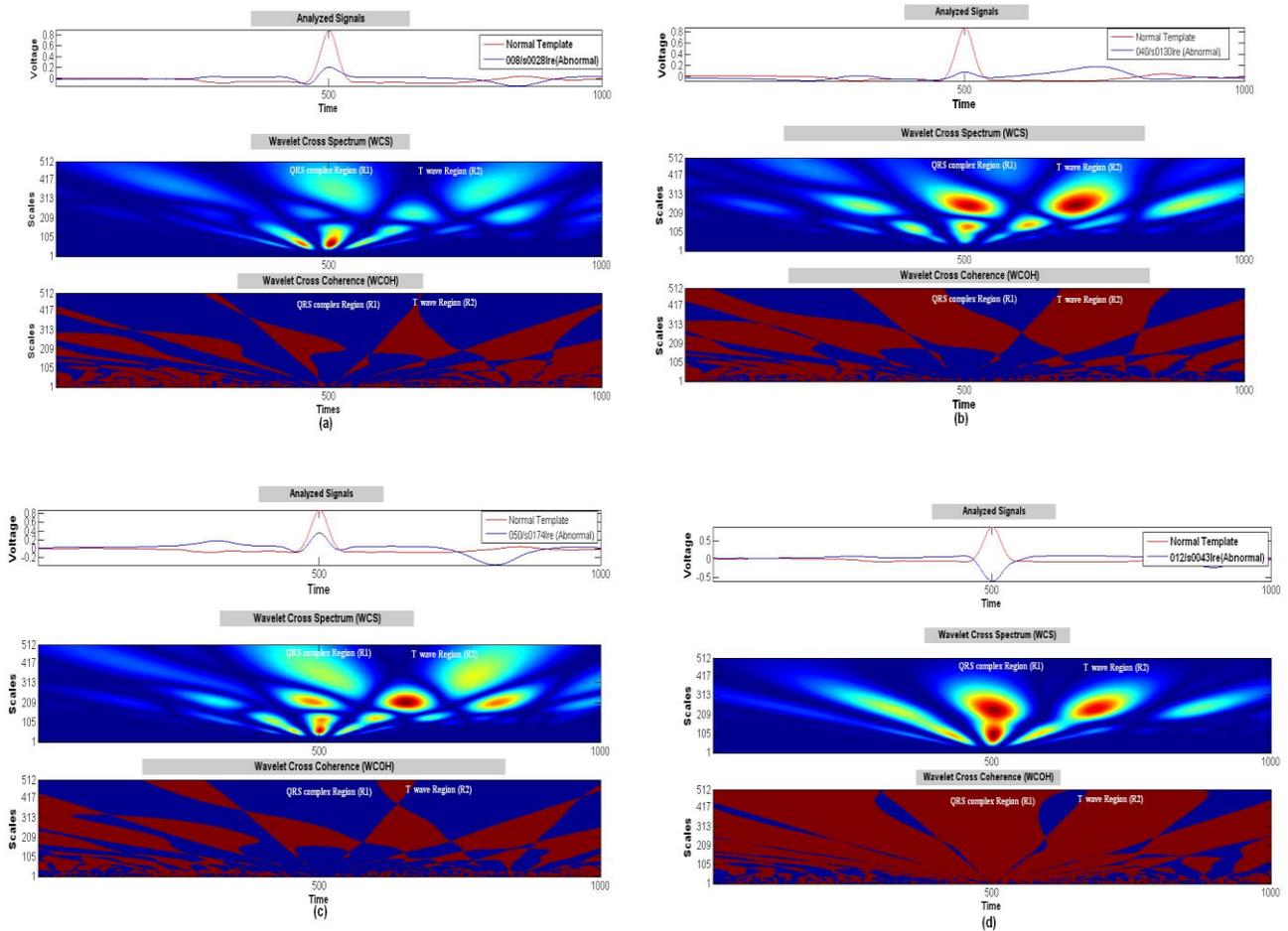

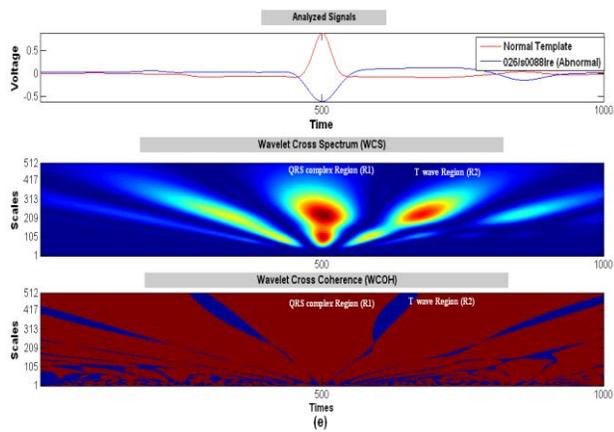
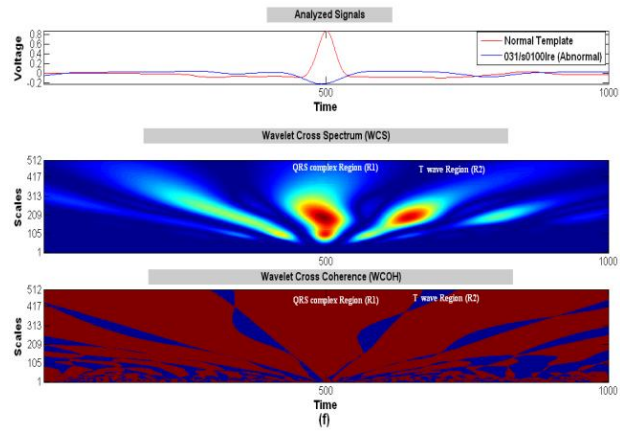

Fig.2: WCS and WC between the standard normal template and abnormal ECG

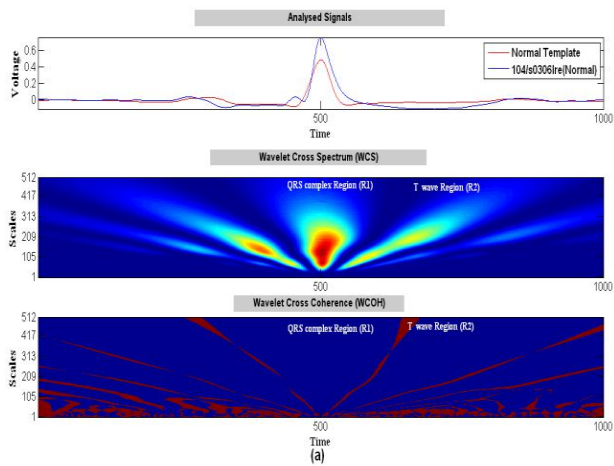
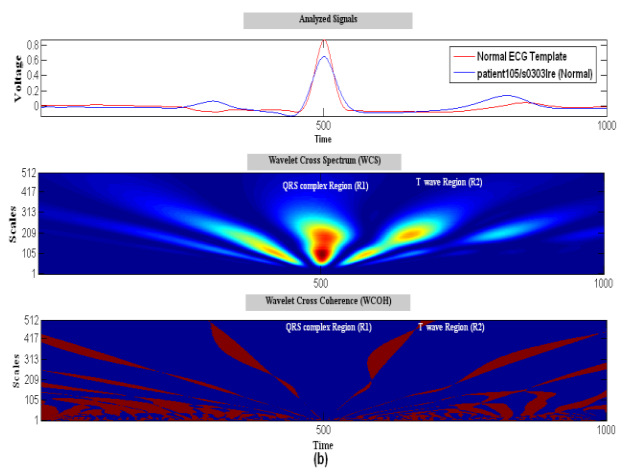
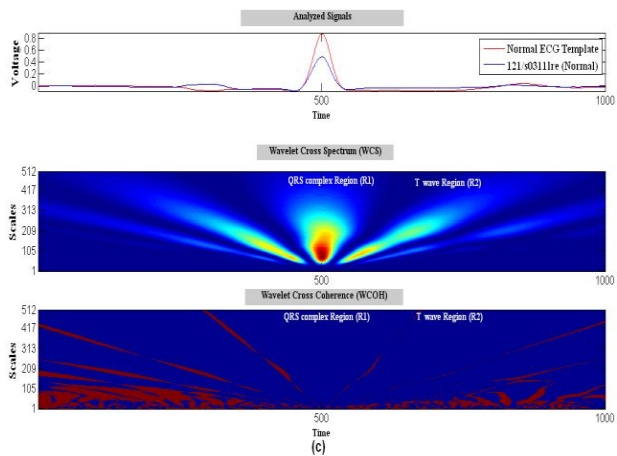
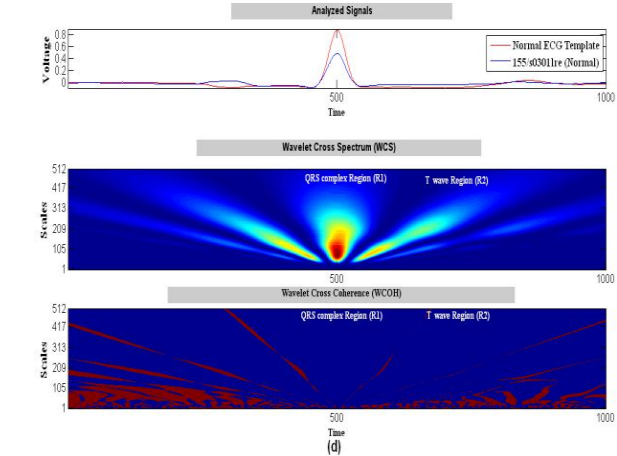

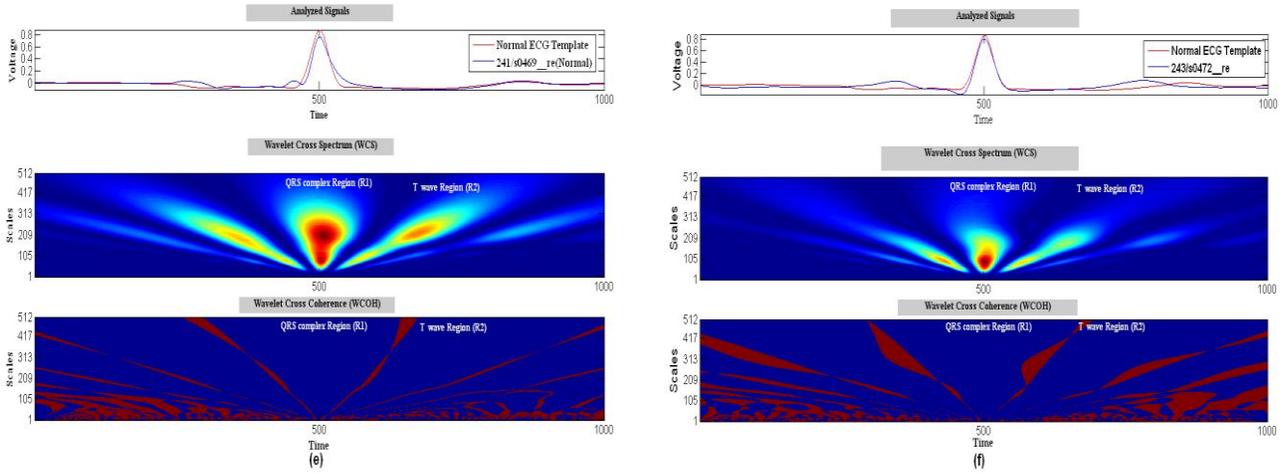

Fig.3: WCS and WC between the standard normal template and a normal ECG data.

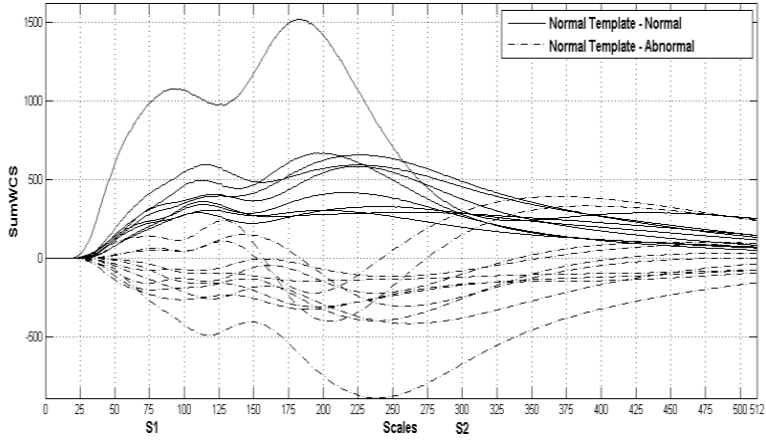

Fig4. Plot for identification of scale range

### 4.3 Performance evaluation Metrics for beat classification over Lead III

The spectrogram of Fig. 2 and Fig. 3 shows the results when applied over pair of normal template-normal and normal template–abnormal beats. Parameter (or feature) extracted from this scale range over the QT zone will produce a unique identification signature.

In this study we have considered four statistical indices: Accuracy ($Acc$), Sensitivity ($Se$), Specificity ($Sp$), which are defined in the following eqns. (6) – (8), respectively.

The most crucial metric for determining overall system performance is usually accuracy. We defined the overall accuracy of the classifier for each file as follows:

$$Acc = \frac{N_T - N_E}{N_T} \times 100 \qquad (6)$$

In this equation, *Acc* is the accuracy, and the variables, $N_E$ and $N_T$, represent the total number of classification errors and beats in the file, respectively.

Sensitivity, *Se*, the ratio of the number of correctly detected events, *TP* (true positives), to the total number of events is given by:

$$Se = \frac{TP}{TP + FN} \times 100 \qquad (7)$$

Where, *FN* (false negatives) is the number of missed events.

The specificity, *Sp*, the ratio of the number of correctly rejected nonevents, *TN* (true negatives), to the total number of nonevents is given by:

$$Sp = \frac{TN}{TN + FP} \times 100 \qquad (8)$$

Where, *FP* (false positives) is the number of falsely detected events. The normal and IMI classification results are stated in the Table I. Where accuracy for lead III is 99.43% and the Se and Sp values are 98.83% and 98.80% respectively in Table 2.

Table I

Performance Evaluation Metric For beat Classification over Lead III

| TP | TN | FP | FN | Se(%) | Sp(%) | Acc.(%) |
|---|---|---|---|---|---|---|
| 2806 | 15282 | 338 | 63 | 98.83 | 98.80 | 99.43 |

Once the abnormal class is differentiated from the normal class k-nn classification rule, with k=3 is applied over the abnormal data set to further distinguish between the Type 1 and Type 2 category. The classification result is stated in Table 2. Figure 5 depicts the pictorial description of the classification of the data set.

Table 2

Classification Accuracy for Type 1 and Type 2 IMI.

| Metric\Class Type | Type 1 | Type 2 |
|---|---|---|
| No. Of Data | 926 | 1880 |
| Correct Classification | 820 | 1636 |
| Misclassification | 106 | 244 |
| Accuracy | 88.5 | 87.02 |

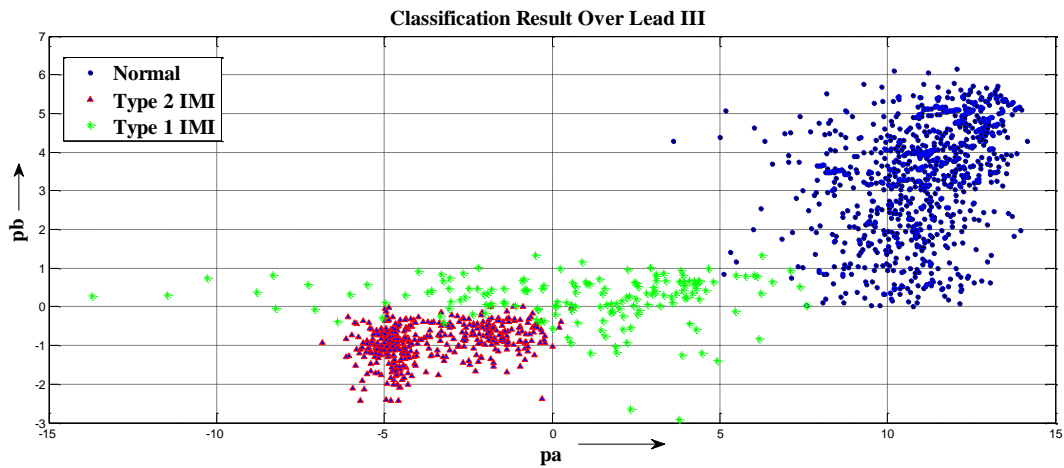

Fig. 5: Classification results

## 5. CONCLUSION

This paper presents a method for analysis of ECG patterns using Cross Wavelet Transform (XWT) and Wavelet Coherence (WCOH) techniques. The application of the Continuous Wavelet Transform to two time series and the cross examination of the two decomposition reveals localized similarities in time and scale. Morlet wavelet is used as the mother wavelet. From, the analysis it was found that wavelet cross spectrum and wavelet coherence reveals great insight into the dissimilarities of the data over which it works. Region based differences are visible in WCS and WCOH of normal-normal and normal-abnormal pairs. Visible dissimilarities in the regions R1 and R2 are marked and after selection of scale appropriate classification parameters are established empirically. Parameters are identified for classification of Type 1 IMI (non Q type, with ST elevation and attenuated QRS complex) and Type 2 IMI (Q type MI with deep Q and inverted T) and normal subjects. Accuracy of 99.43% is obtained for normal and abnormal class categorization. It is observed as 88.5% and 87.02% for Type I and Type 2 MI class respectively. This revealed result further opens scopes for extending this developed method for 12-lead ECG screening systems and addressing other cardiac abnormality issues.

**Acknowledgement:**

The authors are thankful for the grant received from TEQIP Phase II, Department of applied Physics, University of Calcutta.